# Sparsity Aware Normalized Least Mean p-power Algorithms with Correntropy Induced Metric Penalty

Wentao Ma,Hua Qu,Jihong Zhao,Badong Chen
School of Electronic and Information Engineering
Xi'an Jiaotong University
Xi'an, China
E-mail: chenbd@mail.xjtu.edu.cn

Guan Gui
Dept. of Electronics and Information Systems
Akita Prefectural University
Yurihonjo, Japan
E-mail:guiguan@akita-pu.ac.jp

*Abstract*— **For identifying the non-Gaussian impulsive noise systems, normalized LMP (NLMP) has been proposed to combat impulsive-inducing instability. However, the standard algorithm is without considering the inherent sparse structure distribution of unknown system. To exploit sparsity as well as to mitigate the impulsive noise, this paper proposes a sparse NLMP algorithm, i.e., Correntropy Induced Metric (CIM) constraint based NLMP (CIMNLMP). Based on the first proposed algorithm, moreover, we propose an improved CIM constraint variable regularized NLMP(CIMVRNLMP) algorithm by utilizing variable regularized parameter(VRP) selection method which can further adjust convergence speed and steady-state error. Numerical simulations are given to confirm the proposed algorithms.**

## I. Introduction

Based on the assumption of Gaussian noise model, sparsely-aware least mean square(LMS) filtering algorithms(e.g., zero-attracting LMS[1], reweighted zero-attracting LMS[1], sparse regularized least square (RLS) [2], $\ell_0$-norm constrained LMS (L0-LMS) [3], $\ell_p$-norm constrained LMS(LP-LMS) [4] and its variations [5-7]) have been developed and also applied successfully in many applications such as multipath fading channels estimation, underwater acoustic (UWA) channelprobing as well as echo cancellation[8-9]. It is well known that Gaussian assumption has been broadly accepted because of the Central Limit Theorem (CLT). However, this assumption may limit realistic system identification performance, such as non-Gaussian impulsive noise systems.Indeed, many physical experiments have confirmed that impulsive noises are often occurred in many systems such asman-made low frequency atmospheric noise systems and underwater acoustic systems [10-13].By using second-order statistical (SOS) based error criterion, in such situations, sparsity-aware LMS algorithms using may lead to poor performance or instability problem especially in the case ofstronger impulsive interferences.

To solve this problem, some robust adaptive filtering algorithms [14-17] have been proposed. In [14], the least mean p-power (LMP) algorithm was proposed using the p-power of error criterion as the adaptation cost. It is robust to non-Gaussian noise when $p<2$. In the case of AR alpha-stable process, normalized least mean p-norm(NLMP) algorithm using fractional lower order statistics(FLOS)cri-stable terion and the normalized least mean absolute deviation(NLMAD) were proposed in [15-16]. In addition, robust sparsity aware LMP algorithm was also proposed in [17]. Motivated by the previous research, in this paper, we first propose an improved sparse NLMP-type algorithm by adopting sparse constraint function. Since NLMP-type algorithms are still implemented by dividing the step-size parameter by the squared norm of the input vector plus a small regularization parameter. However, the fixed parameter may result in conflict between fast convergence and low mis-adjustment of the NLMP-type algorithms. Motivated by the variable regularized NLMS-type algorithms [18], we propose a variable regularized NLMP (VRNLMP) with sparsity-aware constraint.

Since more efficient sparsity penalty terms (SPT) can enable the filters to take advantage of more sparse prior information in systems, choosing SPT is one of important componentsfor our proposed algorithms. Recent years, the Correntropy Induced Metric (CIM) has been proposed as anexcellent approximation of the $\ell_0$-norm [19, 22].Hence, CIM will be adopted in our proposed algorithms, which are termed as CIMNLMP and CIMVRNLMP, respectively. Simulations results are provided to show the desired performance in convergence, tracking, and mis-adjustment.

The rest of the paper is organized as follows. In Section II, the standard NLMP algorithm is briefly reviewed. In Section III, sparsity aware NLMP algorithms, i.e.,CIMNLMP and CIMVRNLMP, are derived. In Section IV, simulation results are provided to verify the proposed algorithms. Finally, conclusion is given in section V.

This work was supported by National Natural Science Foundation of China (61371807, 61372152).

## II. REVIEW OF NLMP ALGORITHM

We consider the case of the sparse system parameter estimation model as (6). The input vector $\boldsymbol{x}(n) = [x_n, x_{n-1}, \cdots, x_{n-L+1}]^T$ is sent over the FIR system with parameter vector $\boldsymbol{w}^* = [w_1^*, w_2^*, \cdots, w_L^*]^T$, where $L$ is the size of the channel memory. It is assumed that the channel parameters are real-valued, and most of them are zero. The received signal $d(n)$ is modeled as:

$$d(n) = \boldsymbol{w}^{*T}\boldsymbol{x}(n) + v(n), \quad (1)$$

where $v(n)$ is the additive background noise that is independent of $\boldsymbol{x}(n)$.

In many practical situations, the noise $v(n)$ is non-Gaussian due to the impulsive nature of man-made electromagnetic interference as well as nature noises. We will develop in the following sparse regularized NLMP algorithms to address the case under impulsive noise environments.

The least mean $p$-norm (LMP) algorithm is proposed in [14]. It is a generalization of instantaneous gradient decent algorithm, where the gradient of the p-norm of the error is

$$J_{LMP}(n) = |e(n)|^p = |d(n) - y(n)|^p, \quad (2)$$

where $y(n) = \boldsymbol{w}^T(n)\boldsymbol{x}(n)$ is the estimate output signal, $e(n)$ denotes the instantaneous estimation error, $\boldsymbol{w}(n) = [w_1(n), w_2(n), \cdots, w_L(n)]^T$ stands for the estimate of the parameter vector, and the $p$ is a positive constant value. Furthermore, we can keep the filter stable under impulsive noise conditions when c. The gradient descent methods can be used to estimate the filter weights, and an iteration equation can be derived as follows:

$$\begin{aligned}\boldsymbol{w}(n+1) &= \boldsymbol{w}(n) - \eta \frac{\partial J_{LMP}(n)}{\partial \boldsymbol{w}(n)}\\ &= \boldsymbol{w}(n) - \eta\left[-p|e(n)|^{p-1} sgn(e(n))\boldsymbol{x}(n)\right] \quad (3)\\ &= \boldsymbol{w}(n) + \mu|e(n)|^{p-1} sgn(e(n))\boldsymbol{x}(n)\end{aligned}$$

where $\mu = \eta p$ denotes step size. The NLMP algorithm with the motivation of the normalized-LMS algorithm is proposed in [15-16], which uses the following update:

$$\boldsymbol{w}(n+1) = \boldsymbol{w}(n) + \mu \frac{|e(n)|^{p-1} sgn(e(n))}{\|\boldsymbol{x}(n)\|^p + \varepsilon}\boldsymbol{x}(n) \quad (4)$$

where $\varepsilon > 0$ denotes a small positive parameter.

**Remark:** In (4), normalization is obtained by dividing the update term by the $p$-norm of the input vector. The regularization parameter is used to avoid excessively large updates in case of an occasionally small input. Since the overall effective step-size affects the performance of the NLMP, the regularization parameter has an effect on the convergence properties and mis-adjustment as well. In addition, Note that NLMP reduces to the NLMS algorithm when p=2 [18].

## III. SPARSE AWARE NLMP ALGORITHMS

### A. Proposed algorithm: CIMNLMP

The first proposed algorithm is CIMNLMP by incorporating CIM constraint into standard NLMP algorithm. Correntropy is a new local similarity method proposed in [19]. Moreover, some advantage results of the Correntropy are researched in [20-21] further. In particularly, the nonlinear metric-CIM is induced by Correntropy, and it can approximate $l_0$-norm CIM with an appropriately selected kernel width [19, 22]. Now, we give the approximation of the $l_0$-norm based on the CIM. Given a vector $\boldsymbol{x} = [x_1, \cdots, x_N]^T$, and the $\ell_0$-norm can be approximated by CIM with the Gaussian kernel as

$$\|\boldsymbol{x}\|_0 \sim CIM^2(\boldsymbol{x}, 0) = \frac{\kappa(0)}{N}\sum_{i=1}^{N}\left(1 - \exp\left(-\frac{x_i^2}{2\sigma^2}\right)\right) \quad (5)$$

where $\kappa(0) = 1/\sigma\sqrt{2\pi}$. It can be shown that if $|x_i| > \delta$, $\forall x_i \neq 0$, then as $\sigma \to 0$, one can get a solution arbitrarily close to that of the $\ell_0$-norm, where $\delta$ is a small positive number. For CIM minimization, the constrained gradient projection method can be applied. We compute the gradient vector

$$[\boldsymbol{G}] = \left[\frac{\partial CIM^2(\boldsymbol{x}, 0)}{\partial x_1}, \frac{\partial CIM^2(\boldsymbol{x}, 0)}{\partial x_2}, \cdots \frac{\partial CIM^2(\boldsymbol{x}, 0)}{\partial x_N}\right] \quad (6)$$

where

$$\frac{\partial CIM^2(\boldsymbol{x}, 0)}{\partial x_i} = \frac{\kappa(0)}{N\sigma^2} x_i \exp\left(-\frac{x_i^2}{2\sigma^2}\right) \quad (7)$$

We incorporate the gradient of the CIM into the iteration (4) directly, and then the sparse NLMP algorithm can be derived as

$$\begin{aligned}\boldsymbol{w}(n+1) &= \boldsymbol{w}(n) + \mu \frac{|e(n)|^{p-1} sgn(e(n))}{\|\boldsymbol{x}(n)\|^p + \varepsilon}\boldsymbol{x}(n)\\ &- \rho \frac{1}{L\sigma^3\sqrt{2\pi}} \boldsymbol{w}(n) \exp\left(-\boldsymbol{w}^2(n)/2\sigma^2\right)\end{aligned} \quad (8)$$

where the kernel width $\sigma$ is a free parameter in the penalty term. A proper kernel width will make CIM a good approximation to the $\ell_0$-norm constraint [19-20]. The above algorithm is referred to as the CIMNLMP algorithm.

### B. Proposed algorithm: CIMVRNLMP

The role of $\varepsilon$ is to prevent the NLMP-type algorithms [15-16] associated denominator from getting too close to zero, in order to keep the filter away from divergence. However, a too small $\varepsilon$ may results in the denominator very close to zero while a too big $\varepsilon$ may lead to slow convergence speed. In this paper, we propose a sparse NLMP using a variable regularization parameter method and CIM penalty, namely CIMVRNLMP. The proposed algorithm updates $w(n)$ as follows

$$w(n+1) = w(n) + \mu \frac{|e(n)|^{p-1} sgn(e(n))}{\|x(n)\|^p + \varepsilon(n)} x(n) \\ -\rho \frac{1}{L\sigma^3\sqrt{2\pi}} w(n) \exp\left(-w^2(n)/2\sigma^2\right) \quad (9)$$

where $\varepsilon(n) = 1/\theta(n)$, and the error signal power $\theta(n)$ can be estimated as

$$\theta(n) = (1-\Delta)\theta(n-1) + \Delta |e(n)|^p \quad (10)$$

where the positive constant $\Delta$ is close to one.

**Remark:** On the one hand, when the error signal power $\theta(n)$ gets bigger, the regularization parameter $\theta^{-1}(n)$ becomes smaller, and also the effective step-size becomes relatively larger. Consequently, the proposed filter can make bigger adaptation at this mode. On the other hand, when the estimation error is small, our $\theta^{-1}(n)$ gets larger, and it give smaller effective step-size. At this stage, the adaptive filter (9) makes small adjustment. A good property of our proposed algorithm is that, in practice, neither $\theta^{-1}(n)$ nor $\theta(n)$ gets too close to zero.

## IV. NUMERICAL SIMULATION

In this section, we evaluate the performance of the proposed CIMVRNLMP with respect to the LMP, CIMLMP, NLMP, CIMNLMP and VRNLMP algorithms. We assess the proposed filtering algorithm over time-varying channel parameter estimation case under the impulsive noise interference. All the simulation results are obtained by averaging over 100 independent Monte Carlo (MC) runs and 10000 iterations are run for each MC. The convergence is evaluated by mean square deviation(MSD) which is calculated by

$$MSD(n) = E\left\{\|w^* - w(n)\|^2\right\}, \quad (11)$$

where $E(\cdot)$ denotes expectation operator.

In the following simulation, the alpha-stable distribution is considered to model the impulsive noise，which provides a good model for such heavy-tailed noises [23]. The characteristic function of the alpha-stable process is given by

$$f(t) = \exp\left\{j\delta t - \gamma |t|^\alpha [1 + j\beta \operatorname{sgn}(t) S(t,\alpha)]\right\} \quad (12)$$

in which

$$S(t,\alpha) = \begin{cases} \tan\frac{\alpha\pi}{2} & if\ \alpha \neq 1 \\ \frac{2}{\pi}\log|t| & if\ \alpha = 1 \end{cases} \quad (13)$$

where $\alpha \in (0,2]$ is the characteristic factor, $-\infty < \delta < +\infty$ is the location parameter, $\beta \in [-1,1]$ is the symmetry parameter, and $\gamma > 0$ is the dispersion parameter. The characteristic factor $\alpha$ measures the tail heaviness of the distribution. The smaller $\alpha$ is, the heavier the tail is. The parameters vector of the noise model is defined as $V = (\alpha, \beta, \gamma, \delta)$.

The input signal is assumed to be a white Gaussian process with zero mean and unit variance. Assume that the channel memory size is $L = 30$, and the impulsive response of the time-varying system is illustrated in Fig.1. From

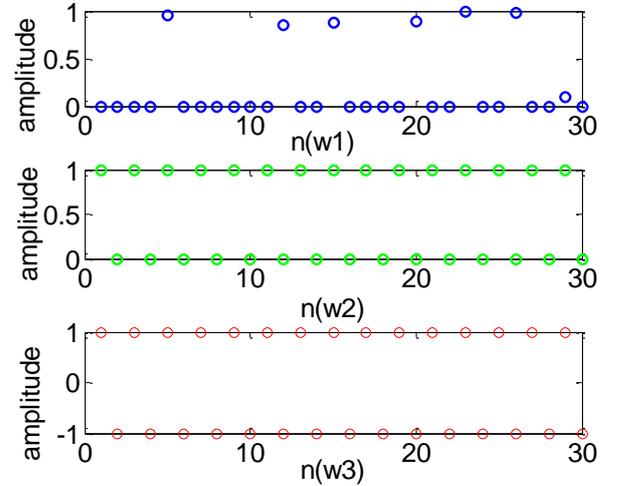

Figure.1 The impulsive response of the time-varying system during three stages

TABLE I. SIMULATION PARAMETERS FOR ALL MENTIONED ALGORITHMS

| Parameter algorithm | $\mu$ | $\rho$ | $\varepsilon$ | $\Delta$ | $\theta(0)$ | $\sigma$ |
|---|---|---|---|---|---|---|
| LMP | 7e-3 | 1e-4 | | | | |
| CIMLMP | 7e-3 | 1e-3 | | | | 1e-2 |
| NLMP | 8e-2 | | 1e-3 | | | |
| CIMNLMP | 9e-2 | 1e-3 | 1e-3 | | | 8e-2 |
| VRNLMP | 8e-2 | | | 0.99 | 1e-5 | |
| CIMVRNLMP | 9e-2 | 1e-3 | | 0.99 | 1e-5 | 8e-2 |

In the first example, we show a comparison among all the proposed algorithms for time-varying channel estimation under the alpha-stable noise environment. All parameters values of the mentioned algorithms are listed in Tab.1.

The p value for all the mentioned algorithms is 1.2. Further, the noise parameter vector is $V = (1.4, 0, 1, 0)$. The convergence in term of the average estimate of the MSD is shown in Fig.2. As one can see, when the system is very sparse (before the 4000th iteration),all mentioned sparse aware algorithms above yield faster convergence rate and better steady-state performances than the LMP algorithm. Note that the CIMVRNLMP achieves the best performance among all the algorithms. After the 4000th iteration, as the number of non-zero taps increase to 15, the performances of the sparse aware algorithms will deteriorate, while CIMVRLMP still maintains the best performance. After 7000 iterations, the CIMNLMP and CIMVRNLMP will perform comparably with the standard LMP even though the system is now completely non-sparse. In the following simulation，we only consider the first stage of the time-varying system due to the strong sparsity.

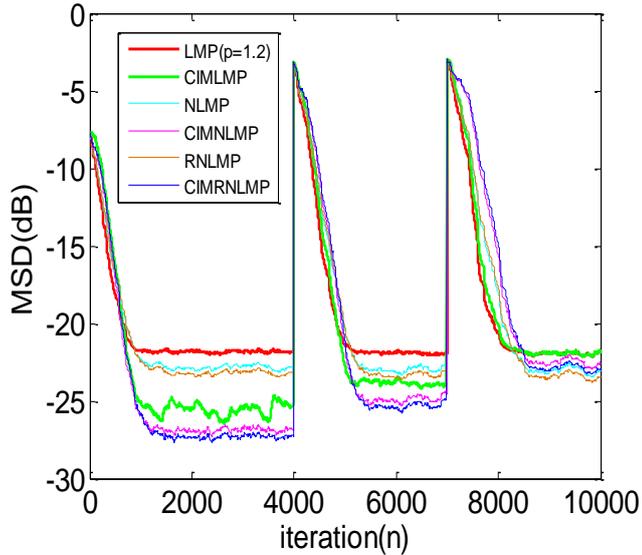

Figure.1.Tracking and steady-state behaviours.

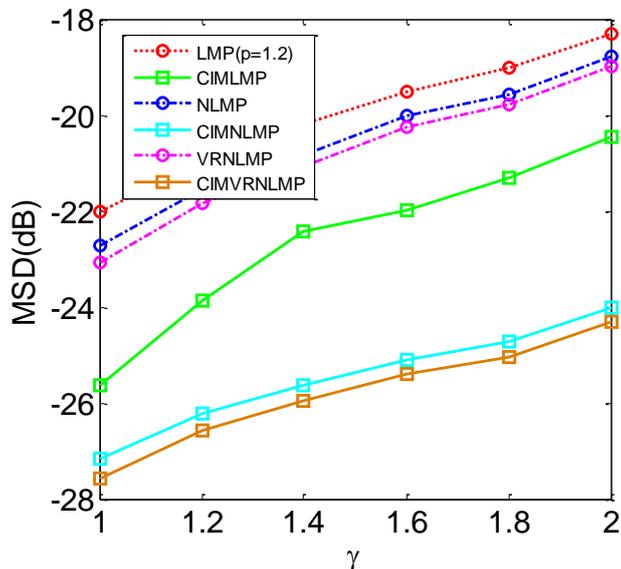

Figure. 3.Steady-state MSD of the channel estimates versus $\gamma$

In the second example, we show the robust property of the sparse LMP algorithms in terms of the dispersion parameter. Note that the parameters are set as the same as the first simulation for all mentioned algorithms above. The ssMSD is calculated as an average of 100 MC simulations. For each MC simulation, the last 1000 iterations are used to compute the ssMSD. Fig.3 illustrates the steady-state MSD versus different $\gamma$ (1 1.2 1.4 1.6 1.8 2), where the parameter $\alpha$ is fixed at 1.4.

Simulation results confirm that the CIMRNLMP may achieve a better accuracy than other algorithms.

In the third example, we investigate the ssMSD of the sparse LMP algorithms as a function of the order p (1, 1.2, 1.4, 1.6, 1.8, and 2) for different $\alpha$ (1, 1.2, 1.4, 1.5, 1.6, 1.8, and 2). The sensitivity of the CIMRNLMP for a range of above parameter values within the algorithm are shown in Fig.4. The parameters are set as $\mu = 0.09$, $\theta(0) = 0.00001$, $\rho = 0.001$, $\Delta = 0.99$, $\sigma = 0.07$ for CIMVRNLMP algorithm in this simulation. We observe that for different p, a more impulsive noise (smaller $\alpha$) leads to a larger MSD. Further, the lowest MSD is obtained when p is close to $\alpha$ .In general, there should be satisfied with $0 < p < \alpha$ between p and $\alpha$ for alpha-stable noise.

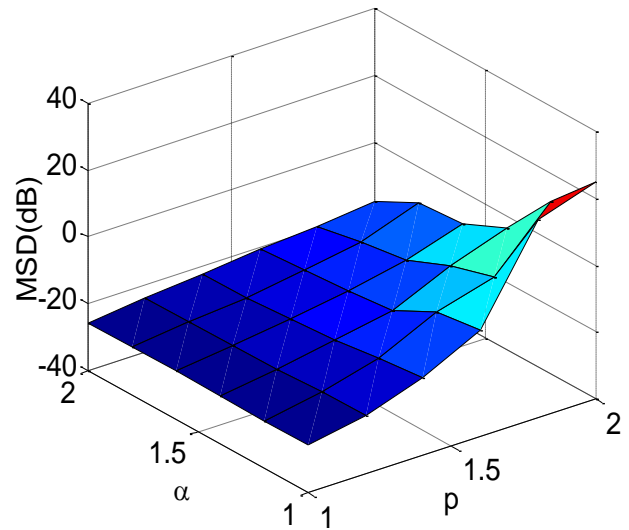

Figure.4 Steady-state MSD of CIMRNLMP for variations in p and $\alpha$

## V.  CONCLUSIONS

We have proposed sparsity aware NLMP algorithms, i.e., CIMNLMP and CIMVRNLMP, which have been shown to perform with fast convergence rate, good tracking, and low misadjustment in impulsive noise environment. Especially，the proposed CIMVRNLMP algorithm uses the variable regularization parameter to enhance the tracking capability of abrupt changes. In addition, the Correntropy induced metric as the sparsity penalty can make the proposed algorithms suitable to sparse system parameter estimation case. Simulation on time-varying system parameter estimation case in impulsive noise environment shows that the proposed method can achieve excellent tracking and steady-state error.